# ZnO/a-Si Distributed Bragg Reflectors for Light Trapping in Thin Film Solar Cells from Visible to Infrared Range

Aqing Chen [1*], Qianmin Yuan [1], Kaigui Zhu [2]

[1] *College of Materials & Environmental Engineering, Hangzhou Dianzi University, Hangzhou 310018, P R China.*

[2] *Department of physics, Beihang University, Beijing 100191, P R China.*

## Abstract

Distributed bragg reflectors (DBRs) consisting of ZnO and amorphous silicon (a-Si) were prepared by magnetron sputtering method for selective light trapping. The quarter-wavelength ZnO/a-Si DBRs with only 6 periods exhibit a peak reflectance of above 99% and have a full width at half maximum that is greater than 347 nm in the range of visible to infrared. The 6-pair reversed quarter-wavelength ZnO/a-Si DBRs also have a peak reflectance of 98%. Combination of the two ZnO/a-Si DBRs leads to a broader stopband from 686 nm to 1354 nm. Using the ZnO/a-Si DBRs as the rear reflector of a-Si thin film solar cells significantly increases the photocurrent in the spectrum range of 400 nm to 1000 nm, in comparison with that of the cells with Al reflector. The obtained results suggest that ZnO/a-Si DBRs are promising reflectors of a-Si thin-film solar cells for light trapping.

## 1. Introduction

Photonic crystals, due to the extraordinary optical properties, are attractive to integrate with solar cells to improve their performance [1,2]. As one dimension photonic crystal, DBRs are the effective reflectors to achieve a desired optical reflectivity by tuning the thickness and refractive index of each layer so that they are integrated with optoelectronic devices including light emitting diodes [3] and surface-emitting lasers[4] and used as rear reflector for light trapping in thin-film solar cells [5–11]. For instance, the DBRs on the back of the InGaN/GaN multiple quantum well solar cells increase the external quantum efficiency in the ultraviolet range by reflecting photons back into the absorber layer [6]. DBRs consisting of ITO and Silica nanoparticle were used as the rear contact of cells [12] to boost the photocurrent of the a-Si thin film cells. However, the efficiency of these DBRs integrated thin-film solar cells did not get remarkable improvement. One of the key reasons for the low conversion efficiency of thin-film solar cells is the weak absorption in the wavelength range of long-wave visible to near-infrared (from 750 to 1500 nm). Intensive investigation has been made to enhance the absorption of a-Si in the this range [2,13,14]. However, most of the methods are to use the expensive noble metal nano-particles with complex techniques [14] for light trapping in infrared range, which are not suitable for industrial products with low cost. Besides, few DBRs with a broad stopband spanning a wide spectrum from near-infrared to visible have been developed to harvest this spectrum for thin-film solar cells.

---

* Corresponding author.
E-mail address: aqchen@hdu.edu.cn

In this work, we demonstrate an economic and simple method to enhance the absorption of a-Si thin film cells which is to utilize the ZnO/a-Si DBRs as the rear reflector for light trapping from visible to infrared range. ZnO thin films which has a low refractive index of 1.9 in the range of visible to near-infrared are widely used in solar cells as the window layer or buffer layer [15–18]. Three kinds of ZnO/a-Si DBRs with a broad stopband from visible to near-infrared range were fabricated by the rf magnetron sputtering method. The first one is quarter-wavelength ZnO/a-Si DBRs (ZnO/a-Si Q-DBRs). The second one is the reversed quarter-wavelength ZnO/a-Si DBRs (ZnO/a-Si R-DBRs) obtained by exchanging the thickness of ZnO and a-Si layer. The third ZnO/a-Si DBRs are the combination of quarter-wavelength and reversed quarter-wavelength ZnO/a-Si DBRs (ZnO/a-Si Q/R-DBRs). Furthermore, Device configurations of a-Si thin films solar cells integrated with the above three kinds of ZnO/a-Si DBRs are modeled for further explore the influence of the ZnO/a-Si DBRs on the performance of a-Si thin films.

## 2. Experiment and Calculation Model

**2.1 Fabrication of ZnO/a-Si DBRs**

The ZnO/a-Si DBRs were deposited on quartz substrates by rf magnetron sputtering. Fig. 1 (a) and

(b) show the image of the ZnO/a-Si Q-DBRs and ZnO/a-Si R-DBRs. Quartz substrates were cleaned ultrasonically with toluene, acetone and ethanol for ten minutes, respectively, and then were put into chamber directly. The ZnO and heavy doped silicon targets were used to deposit the ZnO and heavy doped a-Si layers, respectively. The base pressure of the chamber is less than $5.0 \times 10^{-4}$ Pa. During the deposition, the pressure in the chamber was kept constant at 0.5 Pa and the Argon flow rates were 20 sccm. For ZnO/a-Si Q-DBRs, thicknesses of ZnO and a-Si layer were set around 100 and 56 nm decided by the formula $d_{ZnO} = \lambda/4n_{ZnO}$ and $d_{a-Si} = \lambda/4n_{a-Si}$, respectively. But the thicknesses of ZnO and a-Si layer are 56 nm and 100 nm for ZnO/a-Si R-DBRs, respectively. Both Q-DBRs and R-DBRs have six pairs ZnO/a-Si stack. ZnO/a-Si Q/R-DBRs are the combination of 3-pair quarter-wavelength ZnO/a-Si and 3-pair reversed quarter-wavelength ZnO/a-Si. We used Hitachi S5500 high resolution field emission scanning electron microscopy to characterize the cross section of ZnO/a-Si Q-DBRs, ZnO/a-Si R-DBRs and ZnO/a-Si Q/R-DBRs which are showed clearly in Fig. 1 (c), (d) and (e), respectively.

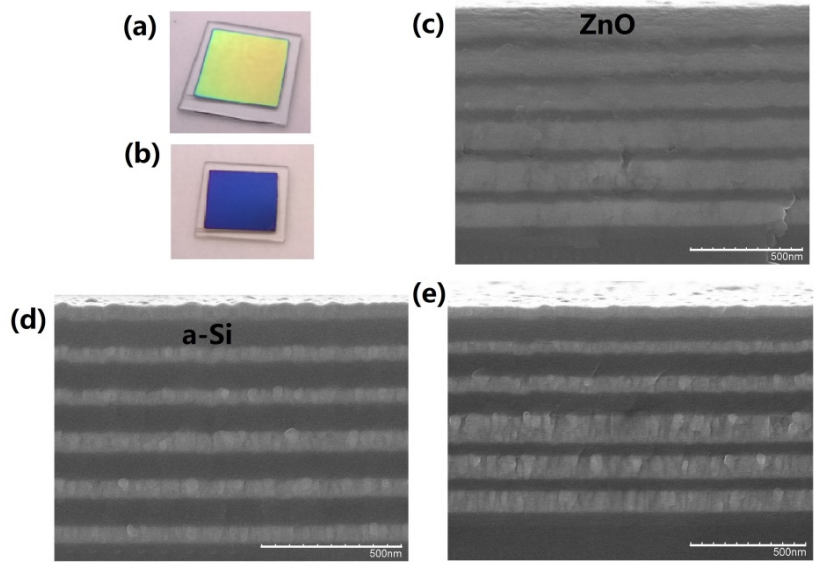

Fig. 1 (a) and (b) are the photograph of ZnO/a-Si Q-DBRs and ZnO/a-Si R-DBRs, respectively. (c) is the SEM image of cross section of the ZnO/a-Si Q-DBRs, (d) is the SEM image of cross section of the ZnO/a-Si R-DBRs, (e) is the SEM image of cross section of the ZnO/a-Si Q/R-DBRs.

## 2.2 Configurations of a-Si thin film solar cells

To investigate the effects of the ZnO/a-Si DBRs on the performance of a-Si thin film solar cells, we build four kinds of models, as shown schematically in Fig. 2. For comparison, a reference case (case 1) of a-Si cells is set up with the aluminum (Al) as the reflector. The second case (case 2) is the a-Si cells with the rear reflector of ZnO/a-Si Q-DBRs. The third one (case 3) is the a-Si cells with the rear reflector of ZnO/a-Si R-DBRs. The last one (case 4) is the a-Si cells with the rear reflector of ZnO/a-Si Q/R-DBRs. The a-Si cells are covered with an anti-reflective coating made of 100 nm transparent conducting oxides. The thickness of ZnO and a-Si in each case is consistent with the experimental value. In the reference case, the thickness of Al is 100 nm.

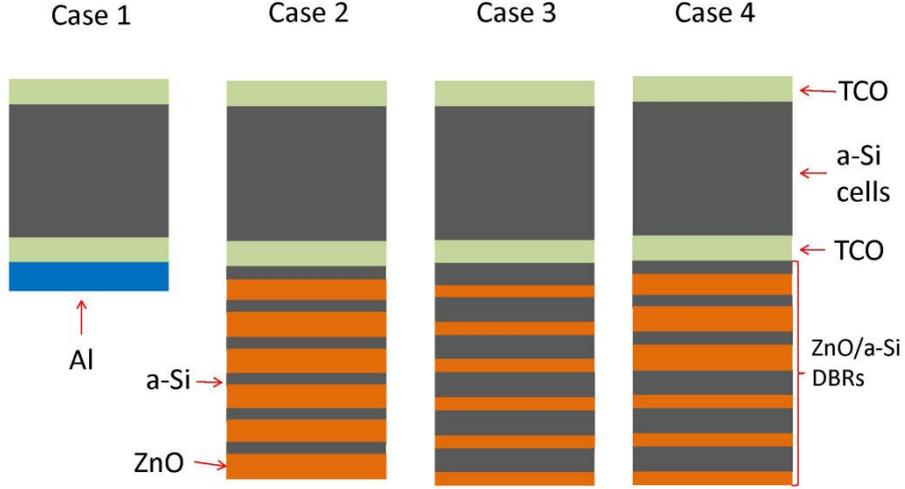

Fig.2 The simulation model of a-Si thin films cells with the rear reflector of Al (case 1), ZnO/a-Si Q-DBRs (case 2), ZnO/a-Si R-DBRs (case 3) and ZnO/a-Si Q/R-DBRs (case 4)

## 3. Results and Analyses

As a result of the high refractive index contrast between ZnO and a-Si, ZnO/a-Si DBRs can achieve a high reflectance and a broad optical forbidden gap easily. The reflectivity of all the samples was measured by a NIR-VIS spectrometer (Cary 5000 UV-Vis-NIR spectrometer) at the incident angle of 3°, as shown in Fig. 3. The peak reflectance of ZnO/a-Si Q-DBRs which have only six quarter-wavelength pairs exceeds 99.1% at 625 nm (the curve 2). ZnO/a-Si R-DBRs also exhibits a high reflectance of above 98.3% at 1050 nm (the curve 1). The stopband widths of ZnO/a-Si Q-DBRs and ZnO/a-Si R-DBRs are 347 nm and 283 nm, respectively. The stopband width characterized by the frequency width $\Delta\omega$ depends on the structure of DBRs. It can be determined by the following gap-midgap ratio[19]

$$\frac{\Delta\omega}{\omega_m} = \frac{4}{\pi} \sin^{-1}\left(\frac{|n_1-n_2|}{n_1+n_2}\right) \qquad (1)$$

where $\omega_m$ is the frequency at the middle gap, $n_1$ and $n_2$ are the refractive index of two layers, respectively. The midgap frequency $\omega_m$ is obtained by

$$\omega_m = \frac{n_1+n_2}{4n_1n_2} \cdot \frac{2\pi c}{a} \tag{2}$$

where *a* is the period constant and *c* is the velocity of light in vacuum. According to above relationship, the quarter-wave structure produces a maximum frequency gap. So, it is obtained that the stopband width of ZnO/a-Si Q-DBRs is larger than that of ZnO/a-Si R-DBRs. It is interesting to note that the ZnO/a-Si Q/R-DBRs exhibits a broad stopband of 668 nm from 686 nm to 1354 nm (from visible to near-infrared), which is described by curve 3 in Fig. 3. This result is similar to other reported reports [20] in which DBRs with n-type a-Si:H/ZnO:Al are modulated to archive a broad stopband. While our results show that the ZnO/a-Si Q/R-DBRs have a broader stopband width which is more than the sum of the stopband width of ZnO/a-Si Q-DBRs and ZnO/a-Si R-DBRs. Nevertheless, there is a dip in the stopband of ZnO/a-Si Q/R-DBRs.

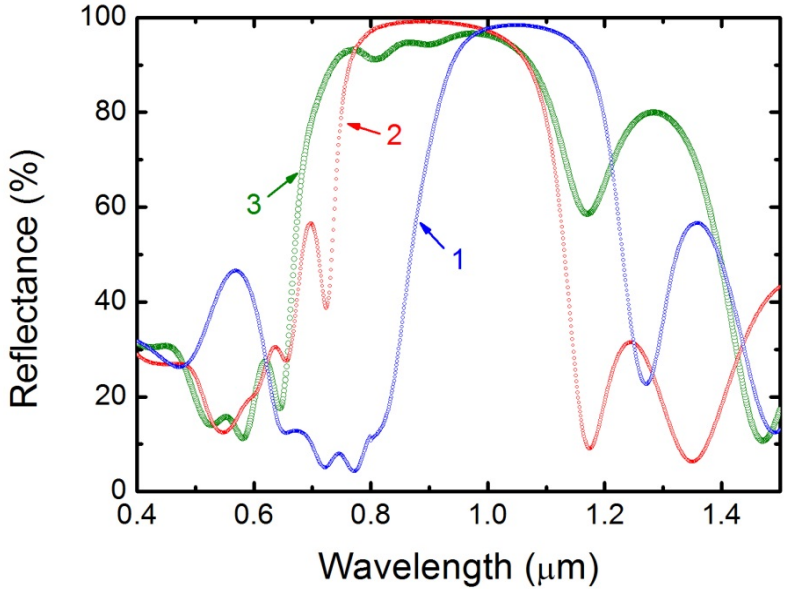

Fig. 3 The reflectivity of ZnO/a-Si R-DBRs, ZnO/a-Si Q-DBRs and ZnO/a-Si Q/R-DBRs, indicating by number 1, 2 and 3, respectively.

The ZnO/a-Si Q/R-DBRs which is the combination of ZnO/a-Si Q-DBRs and ZnO/a-Si R-DBRs should have different reflectance spectra when the incident light impinges from ZnO/a-Si Q-DBRs and R-DBRs, respectively, because ZnO/a-Si Q-DBRs and ZnO/a-Si R-DBRs have different reflection spectrum, as shown in Fig. 3. Fig. 4 shows the reflectivity of ZnO/a-Si Q/R-DBRs when the incident light propagates from ZnO/a-Si Q-DBRs and ZnO/a-Si R-DBRs, respectively. As can be seen in Fig. 4, the stopband width of the ZnO/a-Si Q/R-DBRs is only 400 nm from 954 to 1354 nm when the incident light impinges from ZnO/a-Si R-DBRs. But when the incident light impinges from ZnO/a-Si Q-DBRs, ZnO/a-Si Q/R-DBRs exhibits a broader stopband width of 668 nm from 686 to 1354 nm.

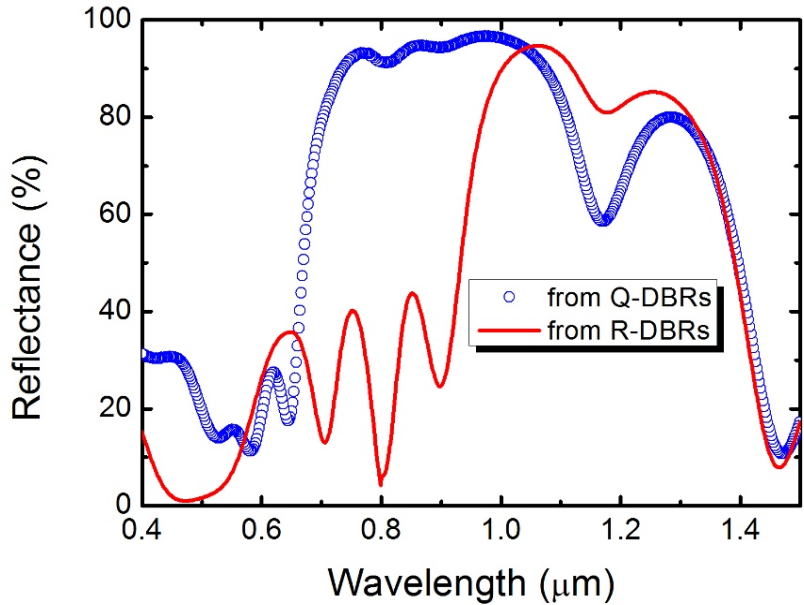

Fig. 4 The reflectivities of ZnO/a-Si Q/R-DBRs at different incident direction, solid line is for light propagation from ZnO/a-Si R-DBRs, dot curve is for light propagation from ZnO/a-Si Q-DBRs.

The distinguished reflectivity of ZnO/a-Si DBRs could play an important role on the light trapping for a-Si thin film solar cells. To evaluate the performance of a-Si thin film solar cells with ZnO/a-Si DBRs, first we have performed calculations to determine the enhancement in optical absorption of the upper a-Si cell with different kinds of reflectors when the thickness of a-Si cells varied from 200 nm to 400 nm. The calculation is carried out by finite difference on time domain (FDTD) method, using a freely available software package [21]. The refractive index and extinction coefficient of amorphous silicon thin film are fitted with the measured complex-refractive index of a-Si by ellipsometry to get parameters for FDTD calculations. The refractive index and extinction coefficient of amorphous silicon thin film are fitted by Lorentz model in the range of 400 nm to 1500 nm, as shown in Fig. 5. It can be seen that the fitted curve agrees well with the experimental curve.

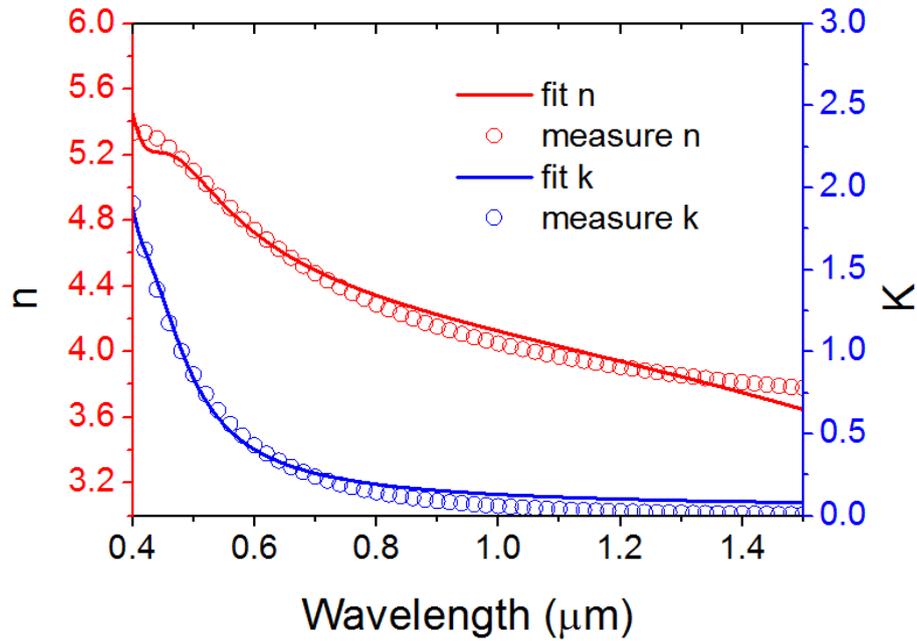

Fig. 5 The refractive index and extinction coefficient of a-Si. Solid line is the measured data and dot curve is the fitted data.

Fig. 6 (a), (b) and (c) show the absorptance spectra of a-Si thin film solar cells with the thickness of 200, 300 and 400 nm, respectively, in all of cases as shown in Fig. 2. It is noteworthy that the a-Si cells with the rear reflector of ZnO/a-Si Q-DBRs and ZnO/a-Si Q/R-DBRs have the same absorptance spectra in the range of 400 nm to 1000 nm in all of the cases. It is observed that the average absorptance of the a-Si cells has been enhanced by ~50% in wavelength range of 700 nm to 900 nm. The 200 nm a-Si cells with the Al reflector have the lowest average absorptance but the 400 nm a-Si cells with the reflector of ZnO/a-Si Q/R-DBRs have the highest average absorptance form 400 nm to 1500 nm. Therefore, it can be obtained that maximum photocurrent is generated in the 400 nm a-Si cells with ZnO/a-Si Q/R-DBRs reflector.

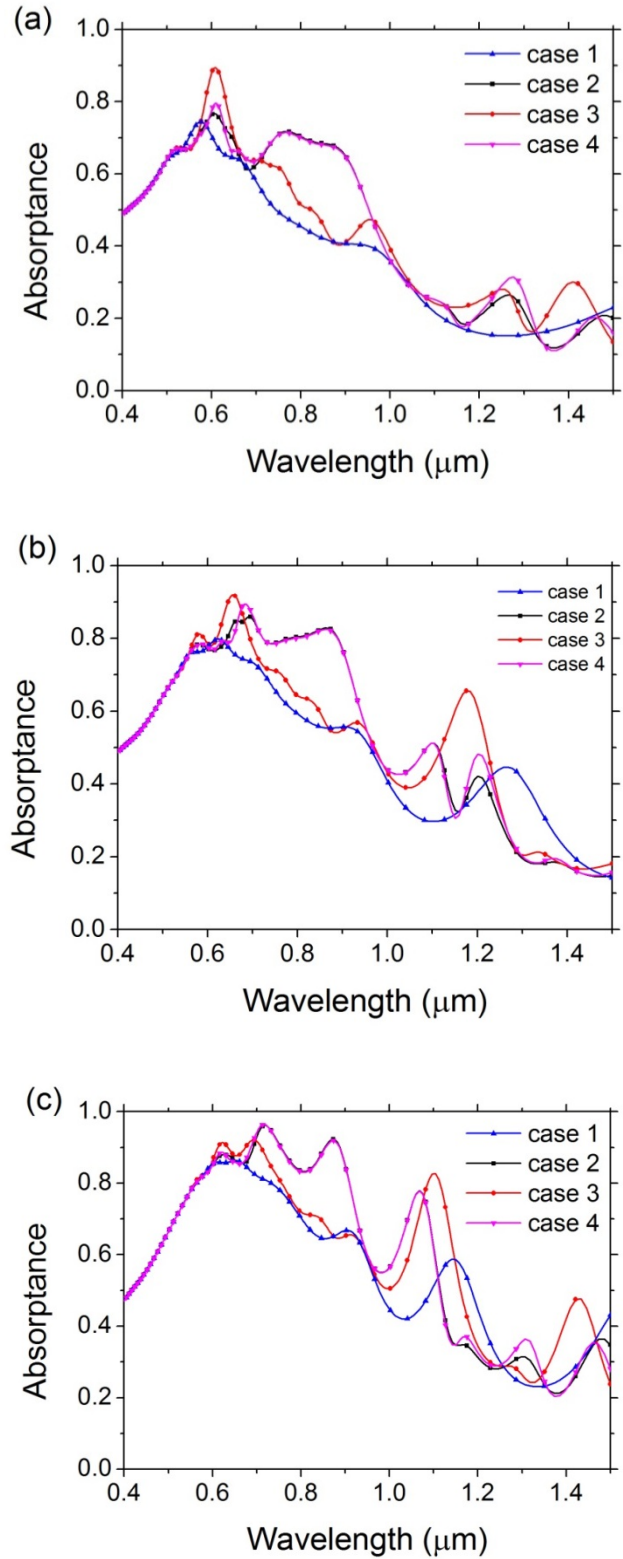

Fig. 6 The absorptance spectra of a-Si cells with the thickness of 200 mn (a), 300 nm (b) and 400 nm (c) when the rear reflector is Al (case1), ZnO/a-Si Q-DBRs (case 2), ZnO/a-Si R-DBRs (case 3) and ZnO/a-Si Q/R-DBRs (case 4)

The photocurrent generated in a-Si cells is greatly dependent on the absorption, upon exposure to the ASTMAM1.5 (Globaltilt) solar spectrum. The light trapping of ZnO/a-Si DBRs enhances effectively the absorptance of a-Si cells, which benefits the photocurrent significantly. The photocurrent $J_{ph}$ generated in a-Si cells is obtained by the following equation [22]

$$J_{ph} = e \int A(\lambda)\, AM1.5(\lambda) d\lambda \qquad (3)$$

Where e is the electron charge, $A(\lambda)$ is the absorption spectrum obtained by FDTD simulation, and the $AM1.5(\lambda)$ is the standard AM1.5G spectrum [23]. Fig. 7 shows the photocurrent in wavelength range of 400 nm to 1000 nm as a function of the thickness of the a-Si cells in all cases which are described in Fig. 2. The photocurrent increases with the thickness of a-Si cells. It can be obvious to note that the ZnO/a-Si DBRs used as the rear reflector significantly improve the photocurrent due to the high reflectivity region from visible to near-infrared. The maximum $J_{ph}$ can reach 29.36 mA/cm$^2$ when the thickness of a-Si cells with ZnO/a-Si Q-DBRs reflector reaches 400 nm. The a-Si cells with Al as the reflector exhibits a low photocurrent since Al has absorption loss in this spectra range. It is also interesting to note that the photocurrent generated in 300 nm a-Si cells with ZnO/a-Si Q-DBRs reflector nearly equates to that generated in 400 a-Si cells with Al reflector. Similarly, the photocurrent generated in 200 nm a-Si cells with ZnO/a-Si Q-DBRs reflector nearly equates to that generated in 300 a-Si cells with Al reflector. The short circuit current densities (Jsc) are determined by the photo-generated carrier collection which is sensitive to the thickness of the absorber layer. Making the absorber layer of the cells thin has a positive effect on carrier collection, which benefits the short circuit current density Jsc and fill factor (FF) [24]. Therefore, the 200 nm and 300 nm a-Si cells with ZnO/a-Si Q-DBRs reflectors have better Jsc and FF than the 300 nm and 400 nm a-Si cells with Al reflector, respectively, due to the same photocurrent.

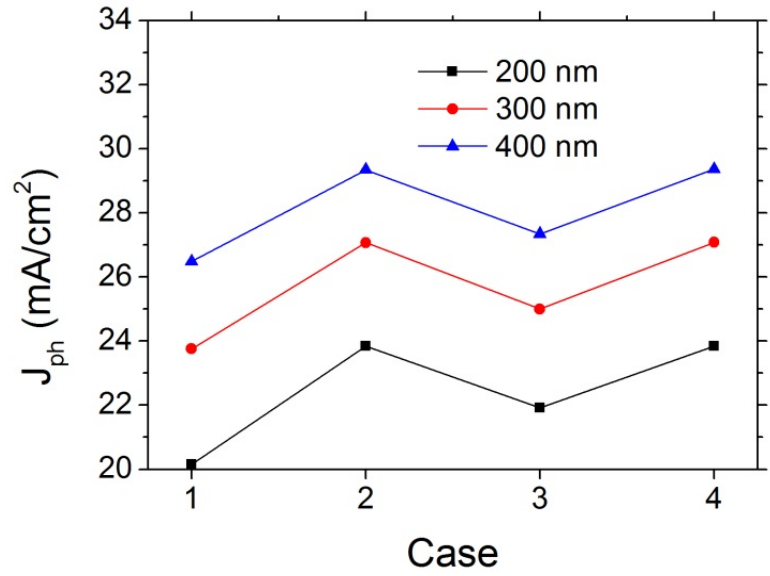

Fig. 7 the photocurrent generated in a-Si cells with different thickness in the wavelength of 400 nm to 1000 nm when the reflector is Al (case1), ZnO/a-Si Q-DBRs (case 2), ZnO/a-Si R-DBRs (case 3) and ZnO/a-Si Q/R-DBRs (case 4).

## 4. Conclusions

In conclusion, we demonstrated the fabrication of ZnO/a-Si DBRs with a peak reflectance of over 99.1% and a broad stopband width from visible to near-infrared region using the magnetron sputtering. ZnO/a-Si Q/R-DBRs composed of ZnO/a-Si Q-DBRs and ZnO/a-Si R-DBRs have wider stopband width from 686 nm to 1354 nm. It is obtained that the ZnO/a-Si DBRs used as the rear reflector of a-Si cells has advantages in trapping the sun light, especially, in the near-infrared range comparing with the conventional Al reflector. Using these ZnO/a-Si DBRs as the rear reflector of a-Si cells effectively enhances the absorption in a-Si cells leading to the improvement of the photocurrent.

# References

[1] H. Li, Q. Tang, F. Cai, X. Hu, H. Lu, Y. Yan, et al., Optimized photonic crystal structure for DSSC, Sol. Energy. 86 (2012) 3430–3437.

[2] H.A. Atwater, A. Polman, Plasmonics for improved photovoltaic devices, Nat. Mater. 9 (2010) 205–213.

[3] H. Guo, X. Zhang, H. Chen, P. Zhang, H. Liu, performance GaN-based LEDs on patterned sapphire substrate with patterned composite SiO 2/Al 2 O 3 passivation layers and TiO 2/Al 2 O 3 DBR backside reflector., Opt. Express. 21 (2013) 1193–1195.

[4] K. Iga, Surface-emitting laser-its birth and generation of new\noptoelectronics field, IEEE J. Sel. Top. Quantum Electron. 6 (2000) 1201–1215.

[5] W.C. Tien, a. K. Chu, ITO distributed Bragg reflectors fabricated at low temperature for light-trapping in thin-film solar cells, Sol. Energy Mater. Sol. Cells. 120 (2014) 18–22.

[6] Y.-L. Tsai, C.-C. Lin, H.-V. Han, C.-K. Chang, H.-C. Chen, K.-J. Chen, et al., Improving efficiency of InGaN/GaN multiple quantum well solar cells using CdS quantum dots and distributed Bragg reflectors, Sol. Energy Mater. Sol. Cells. 117 (2013) 531–536.

[7] X.-L. Zhang, J.-F. Song, X.-B. Li, J. Feng, H.-B. Sun, Anti-reflection resonance in distributed Bragg reflectors-based ultrathin highly absorbing dielectric and its application in solar cells, Appl. Phys. Lett. 102 (2013) 103901.

[8] W. Yu, L. Shen, Y. Long, P. Shen, W. Guo, W. Chen, et al., Highly efficient and high transmittance semitransparent polymer solar cells with one-dimensional photonic crystals as distributed Bragg reflectors, Org. Electron. 15 (2014) 470–477.

[9] M. Zeman, O. Isabella, K. Jäger, R. Santbergen, S. Solntsev, M. Topic, et al., Advanced Light Management Approaches for Thin-Film Silicon Solar Cells, Energy Procedia. 15 (2012) 189–199.

[10] D.B. Bushnell, N.J. Ekins-Daukes, K.W.J. Barnham, J.P. Connolly, J.S. Roberts, G. Hill, et al., Short-circuit current enhancement in Bragg stack multi-quantum-well solar cells for multi-junction space cell applications, Sol. Energy Mater. Sol. Cells. 75 (2003) 299–305.

[11] H. Soh, J. Yoo, D. Kim, Optimal design of the light absorbing layer in thin film silicon solar cells, Sol. Energy. 86 (2012) 2095–2105.

[12] P.G. O'Brien, Y. Yang, A. Chutinan, P. Mahtani, K. Leong, D.P. Puzzo, et al., Selectively transparent and conducting photonic crystal solar spectrum splitters made of alternating sputtered indium-tin oxide and spin-coated silica nanoparticle layers for enhanced photovoltaics, Sol. Energy Mater. Sol. Cells. 102 (2012) 173–183.

[13] M. Faryad, A. Lakhtakia, Enhancement of light absorption efficiency of amorphous-silicon thin-film tandem solar cell due to multiple surface-plasmon-polariton waves in the near-infrared spectral regime*, Opt. Eng. 52 (2013) 087106.

[14] X. Yang, S. Zhou, D. Wang, J. He, J. Zhou, X. Li, et al., Light Trapping Enhancement in a Thin Film with 2D Conformal Periodic Hexagonal Arrays, Nanoscale Res. Lett. 10 (2015) 284.

[15] D. Antartis, I. Chasiotis, Residual stress and mechanical property measurements in amorphous Si photovoltaic thin films, Sol. Energy. 105 (2014) 694–704.

[16]     P. Baviskar, A. Ennaoui, B. Sankapal, Influence of processing parameters on chemically grown ZnO films with low cost Eosin-Y dye towards efficient dye sensitized solar cell, Sol. Energy. 105 (2014) 445–454.

[17]     F. Khan, S.-H. Baek, S.N. Singh, P.K. Singh, M. Husain, J.H. Kim, Influence of Al content on surface passivation properties of Al rich ZnO films for solar cell application, Sol. Energy. 110 (2014) 595–602.

[18]     M. Kriisa, R. Sáez-Araoz, C.-H. Fischer, T. Köhler, E. Kärber, Y. Fu, et al., Study of Zn(O,S) films grown by aerosol assisted chemical vapour deposition and their application as buffer layers in Cu(In,Ga)(S,Se)2 solar cells, Sol. Energy. 115 (2015) 562–568.

[19]     J.D. Joannopoulos, S.G. Johnson, J.N. Winn, R.D. Meade, Photonic Crystals: Molding the Flow of Light, 2nd ed., PRINCETON UNIVERSITY PRESS, PRINCETON AND OXFORD, 2008.

[20]     J. Krc, M. Zeman, S.L. Luxembourg, M. Topic, Modulated photonic-crystal structures as broadband back reflectors in thin-film solar cells, Appl. Phys. Lett. 94 (2009) 6–9.

[21]     A.F. Oskooi, D. Roundy, M. Ibanescu, P. Bermel, J.D. Joannopoulos, S.G. Johnson, Meep: A flexible free-software package for electromagnetic simulations by the FDTD method, Comput. Phys. Commun. 181 (2010) 687–702.

[22]     X. Sheng, S.G. Johnson, L.Z. Broderick, J. Michel, L.C. Kimerling, Integrated photonic structures for light trapping in thin-film Si solar cells, Appl. Phys. Lett. 100 (2012) 111110.

[23]     ASTMG173-03, Standard Tables for Reference Solar Spectral Irradiances: Direct Normal and Hemispherical on 37 degree Tilted Surface (ASTM International, West Conshohocken, Pennsylvania, 2005)

[24]     T. Markvart, L. Castafier, Solar Cells: Materials,Manufacture and Operation, First ed, Elsevier Ltd, Oxford, 2005.